\documentclass[twocolumn, eqsecnum, aps]{revtex4}
\usepackage{graphicx}
\usepackage{dcolumn}

\begin{document}

\title{Relativistic Coupled-Cluster Theory of Atomic Parity Nonconservation:\\
 {\it Application to} $^{137}\text{Ba}^+$}
\vspace*{0.5cm}

\author{$^{1,2}$Bijaya K. Sahoo \protect \footnote[2] {E-mail: bijaya@iiap.res.in, B.K.Sahoo@gsi.de}, $^{1}$Rajat Chaudhuri, $^{1}$B.P. Das,
 $^{3}$Debashis Mukherjee \\
\vspace{0.3cm}
$^{1}${\it Non-Accelerator Particle Physics Group,\\ Indian Institute of 
Astrophysics, Bangalore-34, India}\\
$^{2}${\it Atomphysik, Gesellschaft f\"ur Schwerionenforschung mbH, Germany}\\
$^{3}${\it Department of Physical Chemistry, \\Indian Association for Cultivation of Science, Calcutta-700 032, India}}
\date{Recieved date; Accepted date}

\begin{abstract}
\noindent
We report the result of our {\it ab initio} calculation of the $6s ^2S_{1/2} \rightarrow 5d ^2D_{3/2}$ parity nonconserving electric dipole transition amplitude in $^{137}\text{Ba}^+$ based on relativistic coupled-cluster theory. Considering single, double and partial triple excitations, we have achieved an accuracy of less than one percent. If the accuracy of our calculation can be matched by the proposed parity nonconservation experiment in Ba$^+$ for the above transition, then the combination of the two results would provide an independent non accelerator test of the Standard Model of particle physics.
\end{abstract} 
\maketitle

Parity nonconservation (PNC) in atoms arising from neutral weak
currents has the potential to test the Standard Model (SM) of particle
physics \cite{marciano,ginges}. By combining the results of high precision measurements
and calculations of atomic PNC observables, it is possible to
extract the nuclear weak charge \cite{ginges} and compare with its corresponding value
in the SM. A discrepancy between these two values could
reveal the possible existence of new physics beyond the SM.
The most accurate data on atomic PNC currently comes from the $6s ^2S_{1/2} \rightarrow 7s ^2S_{1/2}$
transition in cesium (Cs), where the claimed experimental \cite{wood} and theoretical \cite{dzuba01} accuracies are 0.35\% and 0.5\% respectively and the deviation from the SM
 is about $1 \sigma$ \cite{dzuba01}. It would indeed be desirable to consider other
candidates which could yield accurate values of the nuclear weak charge.
In this context an experiment to observe PNC in the $6s ^2S_{1/2} \rightarrow 5d ^2D_{3/2}$ transition
in Ba$^+$ using the techniques of ion trapping and laser cooling proposed by Fortson
is of special importance \cite{fortson,fortson02}.

This Letter is concerned with a high precision calculation of the amplitude of the above mentioned parity nonconserving electric dipole (E1$_{PNC}$) transition in Ba$^+$ 
 using relativistic coupled-cluster (RCC) theory, which is equivalent to all order relativistic many-body perturbation theory \cite{lindgren}. It is the first application of this theory to atomic PNC. Blundell {\it et al} had used this theory in the linear approximation to calculate E1$_{PNC}$ for the $6s ^2S_{1/2} \rightarrow 7s ^2S_{1/2}$ transition in Cs \cite{blundell01}. Dzuba {\it et al} \cite{dzuba02} and Geetha \cite{geetha03} have calculated this PNC amplitude for the $6s ^2S_{1/2} \rightarrow 5d ^2D_{3/2}$ transition in Ba$^+$ as discussed later.

  The parity nonconserving nuclear spin independent (NSI) interaction arises from the nucleon-electron neutral weak interaction and its Hamiltonian is given by 
\begin{eqnarray}
\text{H}_{\text{PNC}}^{\text{NSI}} &=& \frac {G_F}{2\sqrt{2}} Q_w \gamma_5 \rho_N(r) 
\end{eqnarray}
\noindent
where $G_F$ is the Fermi constant, $Q_w$ is the nuclear weak charge which is equal to [$(2Z+N)c_{1u} + (2N+Z)c_{1d}$] with $c_{1u}$ and $c_{1d}$ representing electron-up-quark and electron-down-quark coupling constants respectively, $\rho_N(r)$ is the nuclear density function and $\gamma_5(=i\gamma_0\gamma_1\gamma_2\gamma_3)$, which is a pseudo-scalar, is the product of the four Dirac matrices.

$\text{H}_{\text{PNC}}^{\text{NSI}}$ is responsible for mixing atomic states of opposite parities but with the same angular momentum. Its strength is sufficiently weak for it to be considered as a first-order perturbation. It is therefore possible to write the {\it n'th} state atomic wavefunction as
\begin{equation}
|\Psi_n \rangle = |\Psi_n^{(0)} \rangle + G_F |\Psi_n^{(1)} \rangle .
\end{equation}
\noindent
In RCC, the atomic wavefunction $|\Psi_v^{(0)} \rangle$ for a single valence ($v$) open-shell system is given by \cite{lindgren,mukherjee} 
\begin{eqnarray}
|\Psi_v^{(0)} \rangle = e^{T^{(0)}} \{1+S_v^{(0)}\} |\Phi_v \rangle 
\end{eqnarray}
where we define $|\Phi_v \rangle= a_v^{\dagger}|\Phi_0\rangle$, with $|\Phi_0\rangle$ as the Dirac-Fock (DF) state for closed-shell system.

In the singles and doubles approximation we have 
\begin{eqnarray}
T^{(0)} = T_1^{(0)} + T_2^{(0)} , \nonumber \\
S_v^{(0)} =  S_{1v}^{(0)} + S_{2v}^{(0)}  
\end{eqnarray}
\noindent
where $T_1^{(0)}$ and $T_2^{(0)}$ are the single and double particle-hole excitation operators for core electrons and $S_{1v}^{(0)}$ and $S_{2v}^{(0)}$ are the single and double excitation operators for the valence electron respectively. The amplitudes
corresponding to these operators can be determined by solving the relativistic
coupled-cluster singles and doubles equations. A subset of important triple excitations have been considered in the determination of the open shell amplitudes $S_{1v}^{(0)}$ and $S_{2v}^{(0)}$ which is described in \cite{kaldor,geetha01}.

Using eqn. (0.2), the explicit form of E1$_{PNC}$, is given by 
\begin{eqnarray}
\text{E1}_{PNC} = \frac {\langle \Psi_f| \text{D} |\Psi_i \rangle } {\sqrt{\langle \Psi_f|\Psi_f \rangle \langle \Psi_i|\Psi_i \rangle }} \nonumber \\
= \frac {\langle \Psi_f^{(0)}| \text{D} |\Psi_i^{(1)} \rangle + \langle \Psi_f^{(1)}| \text{D} |\Psi_i^{(0)} \rangle } {\sqrt{\langle \Psi_f^{(0)}|\Psi_f^{(0)} \rangle \langle \Psi_i^{(0)}|\Psi_i^{(0)} \rangle }} 
\end{eqnarray}
\noindent
where D is the electric dipole (E1) operator, $i$ and $f$ subscripts are used for initial and final valence electrons respectively. Using the explicit expression for
the first order perturbed wavefunction, we get 
\begin{eqnarray}
\text{E1}_{PNC} = \sum_{I \ne i} \frac {\langle \Psi_f^{(0)}| \text{D} |\Psi_I^{(0)} \rangle \langle \Psi_I^{(0)}| \text{H}_{\text{PNC}}^{\text{NSI}} |\Psi_i^{(0)} \rangle } {E_i - E_I} \nonumber \\
+ \sum_{I \ne f} \frac {\langle \Psi_f^{(0)}| \text{H}_{\text{PNC}}^{\text{NSI}} |\Psi_I^{(0)} \rangle \langle \Psi_I^{(0)}| \text{D} |\Psi_i^{(0)} \rangle } {E_f - E_I}
\end{eqnarray}
where $I$ represent intermediate states.

It is obvious from the above equation that, the accuracy of the calculation of E1$_{PNC}$ depends on the excitation energies of the different intermediate states, the matrix elements of $\text{H}_{\text{PNC}}^{\text{NSI}}$ and D. 
Blundell {\it et al} have used the above equation to determine E1$_{PNC}$ for the 
$6s ^2S_{1/2} \rightarrow 7s ^2S_{1/2}$ transition in Cs by considering the most important intermediate
states \cite{blundell01}. The drawback of this approach is that the summation
can be performed only over a finite set of intermediate states which limits
the accuracy of the calculation. The method we have used in the present work
circumvents this problem by solving the first order perturbed equation
\begin{eqnarray}
(\text{H}^{(0)} - E^{(0)})|\Psi_v^{(1)}\rangle = (E^{(1)} - \text{H}_{\text{PNC}}^{\text{NSI}}) |\Psi_v^{(0)}\rangle .
\end{eqnarray}
where $E^{(1)}$ vanishes for first order correction.

The perturbed cluster operators can be written as
\begin{eqnarray}
T = T^{(0)} + G_F T^{(1)} , \nonumber \\
S_v =  S_v^{(0)} + G_F S_v^{(1)} 
\end{eqnarray}
\noindent
where $T^{(1)}$ and $S_v^{(1)}$ are the first order $G_F$ corrections to the cluster operators  $T^{(0)}$ and $S^{(0)}$ respectively. The amplitudes of these operators are solved, keeping up to liner in PNC perturbed amplitudes, by the following equations
\noindent
\begin{eqnarray}
\langle \Phi_a^p |\overline{\text{H}_N^{(0)}} T^{(1)} + \overline{\text{H}_{\text{PNC}}^{\text{NSI}}}| \Phi_0 \rangle &=& 0 , \nonumber\\
\noindent
\langle \Phi_{ab}^{pq} |\overline{\text{H}_N^{(0)}} T^{(1)} + \overline{\text{H}_{\text{PNC}}^{\text{NSI}}}| \Phi_0 \rangle &=& 0 ,
\end{eqnarray} 
\noindent
and 
\begin{eqnarray}
\langle\Phi_v^p|\overline{\text{H}_N^{(0)}} S_v^{(1)}+(\overline{H_N^{(0)}}T^{(1)}+\overline{\text{H}_{\text{PNC}}^{\text{NSI}}})\{1+S_v^{(0)}\}|\Phi_v\rangle \nonumber \\ =  - \langle\Phi_v^p|S_v^{(1)}|\Phi_v\rangle \text{IP} , \nonumber\\
\langle\Phi_{vb}^{pq}|\overline{\text{H}_N^{(0)}} S_v^{(1)}+(\overline{\text{H}_N^{(0)}}T^{(1)}+\overline{\text{H}_{\text{PNC}}^{\text{NSI}}})\{1+S_v^{(0)}\}|\Phi_v\rangle \nonumber \\ =  - \langle\Phi_{vb}^{pq}|S_v^{(1)}|\Phi_v\rangle \text{IP} ,
\end{eqnarray} 
\noindent
where $\text{H}^{(0)}$ is the Dirac-Coulomb (DC) Hamiltonian and $\overline{\text{H}}$ is defined as $e^{-T^{(0)}}\text{H}e^{T^{(0)}}$ which is computed after determining $T^{(0)}$, IP is the ionization potential energy corresponding to the valence electron '$v$' and the subscript $N$ denotes normal form of an operator. We have used $a,b..$ and $p,q..$ etc. to represent holes and particles respectively. $|\Phi_v^p\rangle$ and  $|\Phi_{vb}^{pq}\rangle$ are the single and double excited states respectively with respect to $|\Phi_v\rangle$. Using Eqns. (0.3), (0.5), (0.8) and only keeping terms linear in $G_F$, the expression for E1$_{PNC}$ can
 be written as
\begin{widetext}
\noindent
\begin{eqnarray}
\text{E1}_{PNC} = \frac {<\Phi_f |\{ 1+ S_f^{(1)^{\dagger}} + T^{(1)^{\dagger}} S_f^{(0)^{\dagger}} + T^{(1)^{\dagger}} \} e^{T^{(0)^{\dagger}}} \text{D} e^{T^{(0)}} \{ 1+ T^{(1)} + T^{(1)} S_i^{(0)} + S_i^{(1)} \} |\Phi_i > }
{\sqrt{(1+N_f^{(0)})(1+N_i^{(0)})}} \nonumber \\
=\frac {<\Phi_f|S_f^{(1)^{\dagger}}\overline{\text{D}^{(0)}}(1+S_i^{(0)})+(1+S_f^{(0)^{\dagger}})\overline{\text{D}^{(0)}}S_i^{(1)}+S_f^{(0)^{\dagger}} (T^{(1)^{\dagger}}\overline{\text{D}^{(0)}}+\overline{\text{D}^{(0)}}T^{(1)}) S_i^{(0)}+(T^{(1)^{\dagger}}\overline{\text{D}^{(0)}}+\overline{\text{D}^{(0)}}T^{(1)})S_i^{(0)}|\Phi_i>}{\sqrt{(1+N_f^{(0)})(1+N_i^{(0)})}} .
\end{eqnarray}
In the above expression we define $\overline{\text{D}^{(0)}} = e^{T^{{(0)}^\dagger}}\text{D}e^{T^{(0)}}$ and $N_v^{(0)} = S_v^{{(0)}^{\dagger}}e^{T^{{(0)}^\dagger}}e^{T^{(0)}}S_v^{(0)}$ for the valence electron '$v$' and each term is connected.
The above matrix element is evaluated by a method similar to that used in our earlier works of Ba$^+$ \cite{geetha02,bijaya01}.
\end{widetext}
The orbitals are constructed as linear combinations of Gaussian type orbitals (GTOs) of the form \cite{rajat02}
\begin{equation}
F_{i,k}(r) = r^k e^{-\alpha_ir^2} .
\end{equation}
where $k=0,1,..$ for s,p,.. type orbital symmetries respectively. For the 
exponents, we have used 
\begin{equation}
\alpha_i = \alpha_0 \beta^{i-1}
\end{equation}

We have considered 30$s_{1/2}$, 25$p_{1/2}$, 25$p_{3/2}$, 25$d_{3/2}$, 25$d_{5/2}$, 20$f_{5/2}$, 20$f_{7/2}$, 20$g_{7/2}$ and 20$g_{9/2}$ GTOs for the DF calculation
and all occupied (active holes) orbitals in the RCC calculations.
We have chosen $\alpha_0$ as 0.00525 and
$\beta$ as 2.73 for all the symmetries. 
All orbitals
are generated on a grid using a two-parameter Fermi nuclear distribution
approximation given  by
\begin{equation}
\rho = \frac {\rho_0} {1 + e^{(r-c)/a}}
\end{equation}
where $\rho_0$ is the average nuclear density, 'c' is the {\it half-charge radius,} and 'a' is related to the {\it skin thickness}.

Our earlier calculations of excitation energies \cite{geetha01}, E1 transition
amplitudes \cite{geetha02} and magnetic dipole hyperfine constants \cite{bijaya01} for some of
the low-lying states in Ba$^+$ based on RCC theory suggest that it is in
principle possible to perform a calculation of E1$_{PNC}$ for the $6s ^2S_{1/2} \rightarrow 5d ^2D_{3/2}$
\begin{table}[t]
\caption{Excitation energy ($cm^{-1}$), E1 transition amplitudes (a.u.) and magnetic dipole hyperfine structure constant (MHz) for different low-lying states of Ba$^+$.} 
\begin{ruledtabular}
\begin{tabular}{l|cccc}
Initial state & $6s ^2S_{1/2}$ & $6s ^2S_{1/2}$ & $5d ^2D_{3/2}$ &
 $5d ^2D_{3/2}$ \\
$\rightarrow$Final state  & $6p ^2P_{1/2}$ & $6p ^2P_{3/2}$ & $6p ^2P_{1/2}$ & $6p ^2P_{3/2}$ \\
\hline
Excitation  &  &  &  & \\
energy & 20410 & 22104 & 15097 & 16795 \\
Expt. \cite{karlsson} & 20262 & 21952 & 15388 & 17079 \\
\hline
E1 transition  &  &  &  & \\
amplitude & 3.37 & 4.72 & 3.08 & 1.36 \\
Expt. \cite{kastberg} & 3.36(0.16) & 4.67(0.08) & 3.03(0.08) & 1.36(0.04) \\
\hline
  &  &  &  & \\
Atomic state & $6s ^2S_{1/2}$ & $6p ^2P_{1/2}$ & $6p ^2P_{3/2}$ & $5d ^2D_{3/2}$ \\
\hline
Hyperfine  &  &  &  & \\
constant (A) & 4078.18 & 740.77 & 128.27 & 189.92 \\
Expt. \cite{blatt,silverans,villemoes} & 4018.871(2) & 743.7(3) & 127.2(2) & 189.7288(6) \\
\end{tabular}
\end{ruledtabular}
\end{table}
transition in that ion to an accuracy of better than one percent. We have
recalculated these quantities using the same method but with a larger basis and
the results are given in table I. The agreement with experiment of the most important excitation
energy ($6p ^2P_{1/2}$) for the calculation of E1$_{PNC}$ is less than one percent.
This is also the case for the hyperfine constants of three of the
states -- $6p ^2P_{1/2}$, $6p ^2P_{3/2}$ and $5d ^2D_{3/2}$, while for the $6s ^2S_{1/2}$ state, the
agreement is a little over one percent. All the transition amplitudes
are within the experimental error bars.
The result of our calculation of the electric quadrupole (E2) amplitude for the $6s ^2S_{1/2} \rightarrow 5d ^2D_{3/2}$ transition is 12.61 in a.u. It is in agreement with our earlier calculation \cite{geetha02} and well within the experimental bounds \cite{yu}. 
In table II, we present the values of the square root of the product
of the hyperfine constants. The accuracies of these two quantities give an indication
of the accuracies of the PNC matrix elements between $6s ^2S_{1/2}$ and $6p ^2P_{1/2}$ states as well as $6p ^2P_{3/2}$ and $5d ^2D_{3/2}$ states. Both of them are in excellent agreement with experiment, suggesting that the two leading PNC matrix elements used in the E1$_{PNC}$ calculation are very accurate.
The contributions from the different terms in E1$_{PNC}$ are presented in
table III. It is clear that the largest contribution comes from $\text{D}S_1^{(1)}$(diagram 1($iii$)) which represents the DF term and a certain sub class of core polarization as well as pair correlation effects \cite{geetha03}. This is due to the relatively large ($6s_{1/2}-6p_{1/2}$) $S_1^{(1)}$ cluster amplitude. Two different types of core polarization effects; $\text{D}T_1^{(1)}$(diagram 1($i$)) and $\text{D}S_2^{(1)}$ as well as its conjugate (diagrams 1($v$) and 1($vi$)) also make significant contributions.The former is mediated by the neutral weak interaction and involves the 6s valence and core electrons. 
Correlation effects corresponding to $S_1^{(0)\dagger}\text{D}S_1^{(1)}$ and  $S_2^{(0)\dagger}\text{D}S_1^{(1)}$ are non negligible, but their signs are opposite. Contributions from other terms are comparatively small.
\begin{table}[t]
\caption{Square root of the magnetic dipole hyperfine constants (MHz) and their deviations from experimental results.}
\begin{ruledtabular}
\begin{tabular}{cccc}
 & Experiment & This work & Deviation (\%) \\
\hline
 & & & \\
$\sqrt{A_{6s ^2S_{1/2}} A_{6p ^2P_{1/2}}}$ & 1728.83 & 1738.1 & 0.5\\
$\sqrt{A_{6p ^2P_{3/2}} A_{5d ^2D_{3/2}}}$ & 155.35 & 156.08 & 0.5 \\
\end{tabular}
\end{ruledtabular}
\end{table}
\begin{table}[t]
\caption{Contributions to the E1$_{PNC}$ calculation in $\times 10^{-11} iea_0 (-Q_W/N)$ using RCC calculation.}
\begin{ruledtabular}
\begin{tabular}{lcclc}
Initial pert. & $6s ^2S_{1/2}^{(1)} \rightarrow $ & & Final pert. & $6s ^2S_{1/2}^{(0)} \rightarrow $ \\
 terms & $ 5d ^2D_{3/2}^{(0)}$&  & terms & $ 5d ^2D_{3/2}^{(1)}$ \\
\hline
 \multicolumn{3}{l}{Dirac-Fock contribution} & & \\
$\text{D} \text{H}_{\text{PNC}}^{\text{NSI}}$ & 2.018 &  & $\text{H}_{\text{PNC}}^{\text{NSI}} \text{D}$ & -0.3 $\times 10^{-5}$ \\
\hline
$\text{D} T_1^{(1)}$ & 0.0003&  & $T^{(1)^{\dagger}} \text{D}$ & 0.418 \\
$\overline{\text{D}^{(0)}} S_{1i}^{(1)} $ & 2.634 & & $ S_{1f}^{(1)\dagger} \overline{\text{D}^{(0)}}$ & -0.179 \\
$\overline{\text{D}^{(0)}} S_{2i}^{(1)}$ & -0.242 & & $S_{2f}^{(1)\dagger} \overline{\text{D}^{(0)}} $ & -0.166 \\
$S_{1f}^{(0)\dagger} \overline{\text{D}^{(0)}} S_{1i}^{(1)}$ & 0.149 & & $ S_{1f}^{(1)\dagger} \overline{\text{D}^{(0)}} S_{1i}^{(0)}$ & 0.003 \\
$S_{1f}^{(0)\dagger} \overline{\text{D}^{(0)}} S_{2i}^{(1)}$ & 0.007  & & $S_{1f}^{(1)\dagger} \overline{\text{D}^{(0)}} S_{2i}^{(0)}$ & 0.008 \\
$S_{2f}^{(0)\dagger} \overline{\text{D}^{(0)}} S_{1i}^{(1)}$ & -0.116 & & $S_{2f}^{(1)\dagger} \overline{\text{D}^{(0)}} S_{1i}^{(0)}$ & -0.009 \\
$S_{2f}^{(0)\dagger} \overline{\text{D}^{(0)}} S_{2i}^{(1)}$ & -0.001 & & $S_{2f}^{(1)\dagger} \overline{\text{D}^{(0)}} S_{2i}^{(0)}$ & 0.001 \\
Norm. & -0.046 & &  & -0.001 \\
 & & & & \\
Total & 2.375 & &  & 0.087 \\
\end{tabular}
\end{ruledtabular}
\end{table}
\begin{figure}[h]
\label{fig:goldstone}
\caption{Important Goldstone diagrams corresponding PNC amplitudes.}
\includegraphics[width=7.5cm]{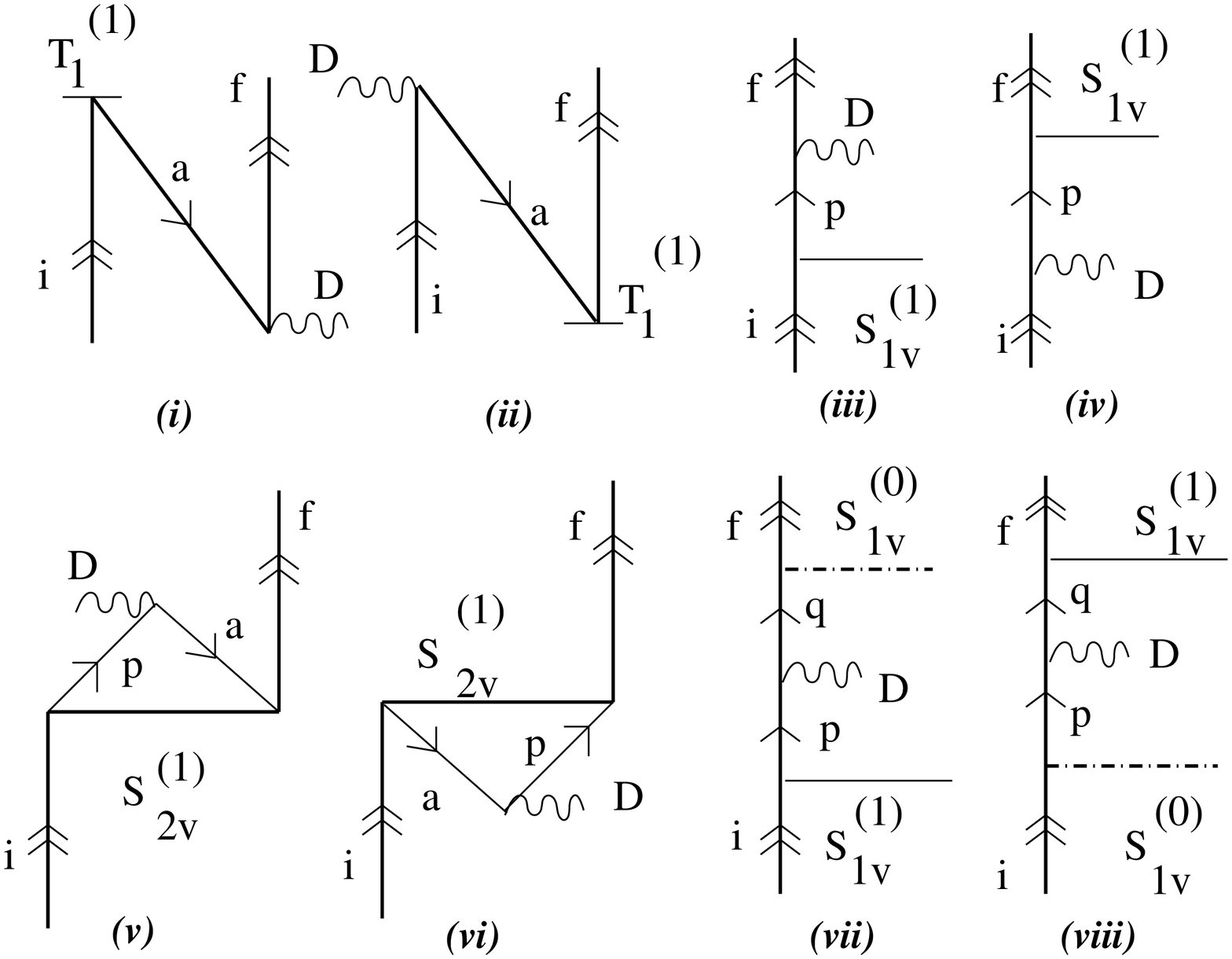}
\end{figure}

The result of E1$_{PNC}$ for the $6s ^2S_{1/2}\rightarrow5d ^2D_{3/2}$ transition in our calculation is 2.46 $\times$ 10$^{-11}iea_0(-Q_W/N)$. It is larger in magnitude than those obtained by Dzuba {\it et al} \cite{dzuba02} and Geetha \cite{geetha03} as shown by table IV. The former work is based on a variant of all order many-body perturbation theory, but it has some semi-empirical features. It is carried out by using two different approaches. One of them is similar to the
sum-over-states approach by Blundell {\it et al} \cite{blundell01} and the other is known as the mixed approach where  the PNC interaction explicitly mixes states of opposite parities. However, both calculations do not include contributions from certain correlation effects; i.e. structural radiation, weak correlation potential and   
normalization of states \cite{dzuba02} that are included in our calculation. Their $6p ^2P_{1/2} \rightarrow 5d ^2D_{3/2}$ E1 matrix element which is important for the above mentioned PNC transition amplitude is not as accurate as ours. Furthermore, the accuracies
of their PNC matrix elements are not known as they have not performed calculations of the hyperfine constants of the relevant states. The reason for the discrepancy between our calculation and Geetha's is that
our approach implicitly includes several intermediate states; particularly doubly excited
opposite parity states which her sum-over-states approach omits. 
\begin{table}[t]
\caption{Comparison of E1$_{PNC}$ results from different calculations in $\times 10^{-11} iea_0 (-Q_W/N)$.}
\begin{ruledtabular}
\begin{tabular}{cc|c|c}
 Dzuba {\it et al} & \cite{dzuba02} & Geetha \cite{geetha03} & Present work\\
 ({\it mixed parity}) & ({\it sum-over-states}) & &  \\
\hline
 2.17 & 2.34  & 2.35 & 2.46 $\pm$ 0.02 \\
\end{tabular}
\end{ruledtabular}
\end{table}

The error accrued in our calculation of E1$_{PNC}$ can be determined from the errors in the excitation energies, E1 transition amplitudes and hyperfine constants (see table I). We have not estimated the errors in the calculated values of these quantities by comparing with measurements, since the error bars in the E1 transition amplitudes are rather large. Instead, we have taken the differences of our RCC calculations with single, double and leading triple excitations and just single and double excitations as the errors. The error in E1$_{PNC}$ (0.02) has been obtained by adding the errors for the different quantities it depends on in quadrature for the leading intermediate states $6p ^2P_{1/2}$ and $6p ^2P_{3/2}$ and using a scale factor to estimate the errors from other intermediate states that together make a small contribution.

The contribution of the Breit interaction to E1$_{PNC}$ at
the DF level is 0.1\% and the nuclear structure contribution
is 0.3\%.The latter has been determined more accurately than
Blundell {\it et al.} \cite{blundell01} using relativistic mean field theory.

In conclusion, we have performed a sub one percent calculation
of E1$_{PNC}$ for the $6s ^2S_{1/2} \rightarrow 5d ^2D_{3/2}$ transition in Ba$^+$
using RCC. We have included single, double as well as a leading
class of triple excitations and highlighted the importance of various
many-body effects. Given the promise that the Ba$^+$ PNC experiment
holds out, it does indeed appear that in the future the result of that
experiment combined with our calculation would constitute a new
and an important probe of physics beyond the SM.
 
We acknowledge discussions with G. Gopakumar. We would like to thank Prof. N. Fortson and J. Sherman for useful communications. We are grateful to Prof. J. Kluge for his critical reading of the paper. BKS thanks DAAD for his scholarship. The calculation was carried out using the Tera-flopp Supercomputer in C-DAC, Bangalore.

\end{document}